\documentclass[aps,prb,twocolumn,showpacs,showkeys]{revtex4}

\usepackage{amsfonts}
\usepackage{amssymb}
\usepackage{amsmath}
\usepackage{array}   
\usepackage{dcolumn} 
\usepackage{bm}      
\usepackage{hyperref}
\usepackage{longtable}
\usepackage{graphicx} 
\usepackage{xcolor}   
\begin{document}


\title{Is reduced-density-matrix functional theory a suitable vehicle
  to import explicit correlations into density-functional
  calculations?}

\author{Peter E. Bl\"ochl}%
\email{peter.bloechl@tu-clausthal.de}%
\author{Christian F. J. Walther}%
\affiliation{Institute for Theoretical Physics, 
Clausthal University of Technology, 
Leibnizstr. 10, 38678 Clausthal-Zellerfeld, Germany}

\author{Thomas Pruschke}%
\affiliation{%
Institut f\"ur Theoretische Physik, Georg-August-Universit\"at
G\"ottingen, Friedrich-Hund-Platz 1, 37077 G\"ottingen, Germany
}

\date{\today}

\begin{abstract}
  A variational formulation for the calculation of interacting fermion
  systems based on the density-matrix functional theory is presented.
  Our formalism provides for a natural integration of explicit
  many-particle effects into standard density-functional-theory based
  calculations and it avoids ambiguities of double-counting terms
  inherent to other approaches.  Like the dynamical mean-field theory,
  we employ a local approximation for explicit correlations.  Aiming
  at the ground state only, trade some of the complexity of Green's
  function based many-particle methods against efficiency. Using short
  Hubbard chains as test systems we demonstrate that the method
  captures ground state properties, such as left-right-correlation,
  beyond those accessible by mean-field theories.
\end{abstract}
\maketitle
\section{Introduction}

One of the most successful theories in solid-state physics is the
density-functional theory (DFT)
\cite{hohenberg64_pr136_864,kohn65_pr140_1133}. Its present
implementations predict properties of materials with an accuracy,
which makes it a standard tool for material scientists both in
fundamental research and in industry.

One of the major deficiencies of currently available density
functionals is the description of materials with strong electron
correlations. Strong electron correlations develop when the
interaction between the electrons dominates over the kinetic
energy. While the kinetic energy favors delocalized electron states
and well-defined energy bands in reciprocal space, strong interactions
may lead to a complete breakdown of the band picture.

Strongly correlated materials often contain 3d transition-metal
elements or rare-earth elements with partially filled f-electron
shells.  The d- and f-electrons are on the one hand localized
core-like states, but, on the other hand, they are located
energetically in the region of valence states. Thus, their nature is
right at the border between localized and delocalized behavior.

Strong correlations are responsible for a wealth of new properties
with potential technological relevance. Among them are heavy-fermion
systems\cite{stewart:1984}, Mott insulators and the Mott-Hubbard
metal-insulator transition (MHMIT)\cite{imada:98}, high-temperature
superconductors\cite{dagotto94_rmp66_763} and the colossal
magnetoresistance
\cite{salamon01_rmp73_583}.  More recently,
the vicinity to phase transitions in those systems has opened the
field of quantum criticality \cite{loehneysen:2007}.

It is a fascinating challenge to unravel the underlying mechanisms
behind these phenomena and eventually to rationally design
applications exploiting them.  A theoretical treatment of these
systems, however, turns out to be a rather big challenge and has been at
the frontier of research in condensed matter physics since nearly half
a century.\cite{imada:98} The use of explicit many-body tools is
nearly impossible for complex materials like perovskites. Therefore,
the many-body community tries to identify the minimal set of degrees
of freedom relevant for the physical properties and sets up a model
hamiltonian for these degrees of freedom. The best known of these
models is the Hubbard model
\cite{hubbard:63,kanamori:63,gutzwiller:63}, which has been the working
horse for correlation effects in transition metal compounds for almost
50 years.

A first attempt to combine DFT with techniques from many-body theory
was made with the GW method \cite{aryasetiawan:98}, which supplements
DFT with certain classes of Feynman diagrams.  
While the GW scheme was applied to semi-conductors with some success
\cite{aryasetiawan:98}, it still fails to describe the physics of the
MHMIT.

A partial success along this line was achieved by the so-called LDA+U
approach, which complements the density functionals by Hartree-Fock
type local correlations\cite{anisimov91_prb44_943}. Hybrid density
functionals\cite{becke93_jcp98_1372}, which replace part of the
exchange energy with the exact non-local exchange energy, capture
essentially the same physics. These functionals lead to a large
improvement of the band gaps, and transition-metal oxides appear
correctly as antiferromagnetic charge-transfer insulators, where
conventional GGA-type functionals erroneously predict metals or a
small d-d band gap.\cite{rollmann04_prb69_165107} It is, however,
important to emphasize that within LDA+U one can only describe
materials with ordered ground states satisfactorily, while the MHMIT
in the paramagnetic state, as in V$_2$O$_3$, cannot be
explained with this method.

A quantum leap in the understanding of the MHMIT was achieved by the
invention of the dynamical mean-field theory (DMFT) \cite{georges:96},
which provides an efficient and reliable tool to approximately solve
the correlation problem for Hubbard type models with local
interactions \cite{imada:98}.  DMFT keeps the local dynamics intact,
which turns out to be essential for the MHMIT. Hence, this theory can
actually describe the competition between the Fermi liquid and the
interaction driven insulator\cite{georges:96,pruschke:95}, and also
allows for the inclusion of ordered phases beyond standard Hartree
theory \cite{obermeier:97a,zitzler:02,pruschke:2005PTP}.  The price to
pay is the neglect of non-local physics and that DMFT fails to capture
phenomena like unconventional superconductivity or quantum criticality
based on spatial fluctuations \cite{loehneysen:2007}. Note, however, that
these effects can be re-included to a certain extent by extending the
original theory \cite{maier:05}

Rather soon it was realized, that the DMFT can be combined with DFT in
a fashion similar to GW.\cite{kotliar:04, Kotliar:2006,
  Held:2006Review} Indeed, early successes of the method comprise the
proper description of spectral properties of $($Sr,Ca$)$VO$_3$, the
MHMIT in V$_2$O$_3$, the energetics of the volume collapse in cerium
\cite{Held:2006Review}, and the lattice properties of plutonium
\cite{savrasov:01}, to name a few.

However, as the DMFT is model-hamiltonian based, it requires the
definition of a Hubbard model prior to its application. This means,
that one needs to extract from DFT (i) the hopping parameters of the
Hubbard model and (ii) the local Coulomb parameters. Both are highly
ill-defined quantities, because the results strongly depend on the basis
set and the sub-manifold of orbitals taken into account. Here,
a large effort has been put forward during the past years to define a
fairly universal interface between the wave-function based DFT and the
Green function based DMFT
\cite{Anisimov:2005,Lechermann:2006,Jacob:2008,Amadon:2008,Haule:2009}.

Besides these technical problems, there are further caveats of the
method.  Firstly, the DFT does already include certain correlation
effects. These have to be ``subtracted'' from the DFT prior to the
calculation of tight-binding and interaction parameters to avoid
double-counting in the DMFT calculation.  This subtle aspect is one of
the most challenging problems in the combination of DFT and DMFT so
far, because nobody knows to what extent and in what form in the
diagrammatic language correlations effects are included in DFT and how
to actually properly 
remove them
 \cite{Kotliar:2006}. 

Secondly, within a many-body formulation, the interactions within the
subspace of the correlated orbitals will be screened by the other
electronic degrees of freedom in the spirit of GW
\cite{Aryasetiawan:2004}. This means, that the actual interactions
become retarded. This poses a serious challenge for the computational
tools necessary to solve the effective quantum impurity model arising
within DMFT.

And last, but not least, the combination of DFT and DMFT cannot be put
rigorously on a variational basis. One can, in principle, formulate an
extremal principle \cite{Anisimov:2005,Kotliar:2006} involving both
wave function for DFT and Green functions for DMFT, but the actual
connection between these two worlds and a proper subtraction of double
counting terms is not obvious.

Therefore, a method, which is fully variational and allows to include
both correlations as they are described by conventional density
functionals and a reasonable and controlled approximation for strong
local correlations, is clearly desirable. The aim of this paper is to
propose such an approach.  The guiding idea is to reformulate the
variational problem in terms of the one-particle density matrix, which
is known as reduced-density-matrix functional theory (rDMFT). This
problem is of the same complexity as a complete many-particle
calculation. Therefore we introduce a local approximation to avoid the
dimensional bottleneck.

The origin of rDMFT goes back to Gilbert's
theorem\cite{gilbert75_prb12_2111}. Gilbert proved that every
ground-state observable can be written as a functional of the
one-particle density matrix. Analogously to the development in the
field of density-functional theory, research is directed towards the
formulation of approximate density-matrix functionals. Most
applications use a variation of M\"uller's
functional\cite{mueller84_pl105A_446, goedecker98_prl81_866,
  csanyi00_prb61_7348, buijse02_molphys100_401,
  gritsenko05_jcp122_204102, sharma08_prb78_201103}.  Recent
functionals have been able to describe the Mott band gap of transition
metal oxides within a non-magnetic
description\cite{sharma08_prb78_201103}, while density functionals
tend to form antiferromagnets instead. This problem is related to the
broken-symmetry states of DFT.

In this paper, we explore the main principles of the method. We do not
yet approach the question on how to exploit the local nature of the
interaction in the Anderson impurity models.

The paper is organized as follows. In the following
section~\ref{sec:theory}, we present the formulation of the theory in
detail. In section~\ref{sec:ci}, we sketch the evaluation of the
reduced-density-matrix functional using a dynamical optimization
similar to the Car-Parrinello method\cite{car85_prl55_2471}.  The
workings of our theory will be explored in section~\ref{sec:model},
where its results for short Hubbard chains are compared with rigorous
many-particle calculations.

\section{Theoretical foundations}
\label{sec:theory}

\subsection{Notation}
While our goal is to incorporate many-particle effects into a general
electronic structure method such as the projector augmented-wave (PAW)
method method\cite{bloechl94_prb50_17953}, the first step will be to
map the electronic structure onto a basisset of local spin orbitals
$\chi_\alpha(\vec{r},\sigma)=\langle\vec{r},\sigma|\chi_\alpha\rangle$.
The spin coordinate $\sigma\in\{\uparrow,\downarrow\}$ can
assume the values spin-up and spin-down. Each orbital is a
two-component spinor, even though we may choose the orbitals such that
they only have one non-zero spin component.

In order to decompose an arbitrary one-particle wave function
$|\psi\rangle$ into the contribution of the orbitals
$|\chi_\alpha\rangle$ according to
\begin{eqnarray}
|\psi\rangle=\sum_\alpha |\chi_\alpha\rangle\langle\pi_\alpha|\psi\rangle
\label{eq:decomposelo}
\,,
\end{eqnarray}
we introduce so-called projector states
$|\pi_\alpha\rangle$. The projector states
$\pi_\alpha(\vec{r},\sigma)=\langle\vec{r},\sigma|\pi_\alpha\rangle$
obey the bi-orthogonality condition
\begin{eqnarray}
\langle\pi_\alpha|\chi_\beta\rangle=\delta_{\alpha,\beta}
\,.
\end{eqnarray}
The identity Eq.~\ref{eq:decomposelo} is valid whenever the wave
function can be expanded completely into the orbital set.  For a
complete orthonormal basisset of local orbitals, the projector states
can also be chosen identical to the local orbitals. However, in the
most general form, the projector functions differ from the orbitals.

The projector states $|\pi_\alpha\rangle$ are conceptually related to
the projector functions of the PAW
method\cite{bloechl94_prb50_17953}. They differ from those used for
the augmentation, because they provide the amplitudes of the local
orbitals instead of partial waves amplitudes provided by the latter.
However, in the PAW method the projector states $|\pi_\alpha\rangle$
for the local orbitals are naturally constructed as superposition of
the projector states for the partial waves.

The creation and annihilation operators in terms of local orbitals can
be expressed by the field operators $\hat{\psi}(\vec{r},\sigma)$ and
$\hat{\psi}^\dagger(\vec{r},\sigma)$ via
\begin{eqnarray}
\hat{c}^\dagger_\alpha&=&\sum_\sigma\int d^3r\;\hat{\psi}^\dagger(\vec{r},\sigma)\pi_\alpha(\vec{r},\sigma)
\\
\hat{c}_\alpha&=&\sum_\sigma\int d^3r\;\pi^*_\alpha(\vec{r},\sigma)\hat{\psi}(\vec{r},\sigma)
\,.
\end{eqnarray}
The back transform is
\begin{eqnarray}
\hat{\psi}^\dagger(\vec{r},\sigma)&=&
\sum_\alpha \chi^*_{\alpha}(\vec{r},\sigma)\hat{c}^\dagger_\alpha
\\
\hat{\psi}(\vec{r},\sigma)&=&
\sum_\alpha \hat{c}_\alpha \chi_{\alpha}(\vec{r},\sigma)
\,.
\end{eqnarray}

Creation and annihilation operators obey the anti-commutator relation
\begin{eqnarray}
\left[\hat{c}^\dagger_\alpha,\hat{c}_\beta\right]_+&=&\langle\pi_\alpha|\pi_\beta\rangle
\\
\left[\hat{c}^\dagger_\alpha,\hat{c}^\dagger_\beta\right]_+&=&0
\\
\left[\hat{c}_\alpha,\hat{c}_\beta\right]_+&=&0\,.
\end{eqnarray}

Note, that the anti-commutator between creators and annihilators,
i.e. the overlap between projector states, is not necessarily equal to
$\delta_{i,j}$. This is due to our generalization to non-orthonormal
local orbitals.

The Hamilton operator is given by 
\begin{eqnarray}
\hat{H}=\hat{h}+\hat{W}\,,
\end{eqnarray}
where $\hat{h}$ is the one-particle part of the Hamilton operator,
containing kinetic energy and external potential. $\hat{W}$ is the Coulomb
repulsion between the electrons.

The non-interacting Hamiltonian 
\begin{eqnarray}
\hat{h}=\sum_{\alpha,\beta}h_{\alpha,\beta}
\hat{c}^\dagger_\alpha\hat{c}_\beta
\end{eqnarray}
has the matrix elements
\begin{eqnarray}
h_{\alpha,\beta}=\langle\chi_\alpha|\frac{\hat{p}^2}{2m_e}+\hat{V}_{ext}
|\chi_\beta\rangle\,.
\end{eqnarray}

The interaction energy has the form
\begin{eqnarray}
\hat{W}=\frac{1}{2}\sum_{\alpha,\beta,\gamma,\delta}
W_{\alpha,\beta,\gamma,\delta}
\hat{c}^\dagger_\alpha\hat{c}^\dagger_\beta
\hat{c}_\delta\hat{c}_\gamma
\label{eq:interactionw}
\end{eqnarray}
with the matrix elements
\begin{eqnarray}
\hspace{-0.25cm}W_{\alpha,\beta,\gamma,\delta}&=&
\sum_{\sigma,\sigma'}\int d^3r\int d^3r'\;
\nonumber\\
&\times&
\frac{e^2\chi^*_\alpha(\vec{r},\sigma)\chi^*_\beta(\vec{r'},\sigma')
\chi_\gamma(\vec{r},\sigma)\chi_\delta(\vec{r'},\sigma')}{4\pi\epsilon_0
|\vec{r}-\vec{r'}|}\,. 
\label{eq:interactionmatrixelements}
\end{eqnarray}
Note the order of the indices of $W_{\alpha,\beta,\gamma,\delta}$ in
Eqs.~\ref{eq:interactionw} and \ref{eq:interactionmatrixelements}.

\subsection{Reduced-density-matrix-functional theory}
As shown by Gilbert\cite{gilbert75_prb12_2111}, every ground-state
property of an electron gas can be expressed as functional of a
one-particle reduced density matrix. Let us review the basis of the
theory:

The ground-state energy for a given external potential can be written
as
\begin{eqnarray}
  E(\mathbf{h})=\min_{|\Phi\rangle,\mathcal{E}}\Bigl(&&
\langle\Phi|\sum_{\alpha,\beta}h_{\alpha,\beta}\hat{c}^\dagger_\alpha\hat{c}_\beta
+\hat{W}|\Phi\rangle
\nonumber\\
&&-\;\mathcal{E}\left(\langle\Phi|\Phi\rangle-1\right)\Bigr)\,.
\label{eq:energyofV}
\end{eqnarray}
The minimization is performed over all fermionic many-particle wave
functions $|\Phi\rangle$. $\mathcal{E}$ is the Lagrange multiplier for
the normalization constraint of the wave function.  The particle
number is not restricted and the Fermi level is placed at the energy
zero: that is, we implicitly employ a grand-canonical description for
the electrons: Electrons are taken from some vacuum level that defines
our energy zero and added to the system as long as this addition
lowers the total energy.

Let us perform a Legendre transform of the total energy,
Eq.~\ref{eq:energyofV}, with respect to the one-particle Hamiltonian
$\mathbf{h}$.
\begin{eqnarray}
F^{\hat{W}}({\bm\rho})&=&
\min_{{\mathbf{h}}}
\left[
 E({\mathbf{h}})-\mathrm{Tr}({\mathbf{h}\bm\rho})\right]\,.
\label{eq:legendretransform}
\end{eqnarray}

The minimum condition of Eq.~\ref{eq:legendretransform}
with respect to $\mathbf{h}$,
\begin{eqnarray}
\rho_{\alpha,\beta}(\mathbf{h})&=&\frac{\partial E}{\partial{h}_{\beta,\alpha}}
=\langle\Phi|\hat{c}^\dagger_\beta\hat{c}_\alpha|\Phi\rangle \;,
\label{eq:defrdm}
\end{eqnarray}
identifies the conjugate variable $\bm{\rho}$ with the
one-particle reduced density matrix of the ground state $|\Phi\rangle$
of the Hamiltonian $\hat{h}+\hat{W}$.

The result of the Legendre transform is the reduced-density-matrix
functional
\begin{eqnarray}
  F^{\hat{W}}({\bm\rho})=\min_{|\Phi\rangle,\mathbf{h},\mathcal{E}}
\Bigl[&&\hspace{-0.3cm}
\langle\Phi|\hat{W}|\Phi\rangle
+\sum_{\alpha,\beta}{h}_{\alpha,\beta}
\left(
\langle\Phi|\hat{c}^\dagger_\alpha\hat{c}_\beta|\Phi\rangle-\rho_{\beta,\alpha}
\right)
\nonumber\\
&&-\;\mathcal{E}\left(\langle\Phi|\Phi\rangle-1\right)\Bigr]
\,.\label{eq:rdmf}
\end{eqnarray}
The reduced-density-matrix functional $F^{\hat{W}}[\bm\rho]$ is
universal in the sense that it is an intrinsic property of the
interacting electron gas and that it is independent of the external
potential.  We introduced the interaction $\hat{W}$ as
superscript in the notation for the reduced-density-matrix functional
because we will use different interactions in the course of this
paper.

The one-particle Hamiltonian $\mathbf{h}$ required in
Eq.~\ref{eq:legendretransform} to produce a specified density matrix
$\bm\rho$ is obtained, up to the different sign, as derivative of the
density-matrix functional with respect to the density matrix, namely as
\begin{eqnarray}
  {h}_{\alpha,\beta}(\bm\rho)
=-\frac{\partial F^{\hat{W}}}{\partial \rho_{\beta,\alpha}}\,.
\end{eqnarray}

The range of allowed arguments $\bm{\rho}$ for the universal
functional is only a subset of all hermitian matrices in the
one-particle Hilbert space. The range is limited to the matrices that
can be constructed by Eq.~\ref{eq:defrdm} with any fermionic
many-particle state $|\Phi\rangle$ from the Fock space. Matrices of
this form are called N-representable\footnote{We use the term
  N-representable in its generalized meaning allowing also density
  matrices obtained from superpositions of states with different
  particle numbers.}

According to a theorem due to Coleman\cite{coleman63_rmp35_668}, the
eigenvalues of all N-representable matrices lie between zero and one
and all matrices with eigenvalues between zero and one are
N-representable.

The eigenstates $|\psi_n\rangle$ of the one-particle
reduced density matrix are called natural
orbitals\cite{loewdin55_pr97_1474} and the eigenvalues $f_n$ are their
occupations. The one-particle density matrix expressed in terms of
natural orbitals and occupations is given by
\begin{eqnarray}
\rho_{\alpha,\beta}=\sum_n\langle\pi_\alpha|\psi_n\rangle f_n
\langle\psi_n|\pi_\beta\rangle\,.
\end{eqnarray}

Hence, the total energy can be expressed by a Legendre back-transform
as
\begin{eqnarray}
E(\mathbf{h})=\hspace{-3mm}\min_{\{|\psi_n\rangle,x_n,\Lambda_{m,n}\}}\Bigl[
\hspace{-4mm}&&
F^{\hat{W}}\left(\left\{
\sum_n\langle\pi_\alpha|\psi_n\rangle f(x_n)\langle\psi_n|\pi_\beta\rangle
\right\}\right)
\nonumber\\&&
+\sum_{n,\alpha,\beta}\langle\pi_\alpha|\psi_n\rangle 
f(x_n)\langle\psi_n|\pi_\beta\rangle{h}_{n,m}
\nonumber\\&&
-\sum_{n,m}\Lambda_{n,m}
\left(\langle\psi_n|\psi_m\rangle-\delta_{m,n}\right)
\Bigr]\,.
\label{eq:energyfromdmf}
\end{eqnarray}
In order to respect the limited range of values for the occupations we
introduce the transformation $f(x)=\sin^2(\frac{\pi}{2}x)$ from an
unrestricted real variable $x$ to the interval $[0,1]$.  The
$\Lambda_{n,m}$ are the Lagrange multipliers for the orthonormality
constraint of the natural orbitals.

Starting from the basic formulation of
reduced-density-matrix-functional theory, the usual route taken is to
find and explore approximate but analytic expressions for the
density-matrix functional.  We follow a different route, namely to
determine the functional on the fly using the explicit
constrained-search method of Levy\cite{levy79_pnas76_6062}.

\subsection{Local approximation\label{subsec:local_approximation}}
Up to now, the formulation has been exact. However, the evaluation of
the density-matrix functional using the constrained search
algorithm\cite{levy79_pnas76_6062}  requires the solution of a full
many-particle problem. Because the many-particle Hilbert space grows
exponentially with system size, we introduce approximations that avoid
this dimensional bottleneck.

To this end, we introduce clusters $\mathcal{C}_R$ of orbitals for
which the correlations will be treated explicitly. Such a cluster may
include all orbitals from the d- or f-electron shell of an atom;
alternatively, it may include all orbitals centered on a single atom,
or it may include all or a certain subset of orbitals from several
atoms. In analogy to the Anderson impurity model, we will refer to
these clusters in the following as "impurities".

Besides the full interaction of Eq.~\ref{eq:interactionw}, we
introduce a local interaction $\hat{W}_{\text{loc}}$. Two electrons interact
with the local interaction only if both reside on the same impurity,
i.e. in the same cluster $\mathcal{C}_R$ of correlated orbitals.
\begin{eqnarray}
\hat{W}_{\text{loc}} &=& \sum\limits_{R}\hat{W}_R
\\
\label{eq:local_interaction}
 \hat{W}_R&:=&\frac{1}{2}\sum_{\alpha,\beta,\gamma,\delta\in \mathcal{C}_R}
W_{\alpha,\beta,\gamma,\delta}\;\hat{c}^\dagger_\alpha
  \hat{c}^\dagger_\beta\hat{c}_\delta\hat{c}_\gamma\,.
\end{eqnarray}
Such an interaction is well-known in model-based studies of interaction
effects. For example, the Hubbard model uses an interaction of this
form.\cite{hubbard:63, gutzwiller:63, kanamori:63}
 
In order to establish the link between DFT and DMFT, let us also
introduce an approximate density-matrix-functional
$F^{\hat{W}}_{DFT}[\rho]$ calculated from a density functional as
\begin{eqnarray}
\hspace{-0.2cm}
F^{\hat{W}}_{DFT}[\bm\rho]:=
\frac{1}{2}\int d^3r\int d^3r'\;
\frac{e^2n(\vec{r})n(\vec{r'})}{4\pi\epsilon_0|\vec{r}-\vec{r'}|}
+E_{xc}[n(\vec{r})]\,,
\label{eq:dftdmf}
\end{eqnarray}
where
$n(\vec{r}):=\sum_{\alpha,\beta}\rho_{\alpha,\beta}\chi^*_{\beta}(\vec{r})
\chi_{\alpha}(\vec{r})$ is the electron density.  

With the approximate functional from Eq.~\ref{eq:dftdmf}, the
density-matrix functional with the full interaction can be written in
the form $
F^{\hat{W}}=F^{\hat{W}}_{DFT}+(F^{\hat{W}}-F^{\hat{W}}_{DFT}) $, i.e.\
as DFT plus a correction describing many-particle effects explicitly.
Irrespective of the choice of the density functional in
Eq.~\ref{eq:dftdmf}, this form of the functional is exact, because it
simply adds and subtracts the same term to and from the full
functional $F^{\hat{W}}$.

The first approximation of our theory is to replace the interaction in
the correction term with the sum of local interactions
\begin{eqnarray}
F^{\hat{W}}\approx F_{DFT}^{\hat{W}}
+\left(F^{\sum_R\hat{W}_R}-F_{DFT}^{\sum_R\hat{W}_R}\right)\,.
\label{eq:approx1a}
\end{eqnarray}
On this level, the theory describes interactions on the impurities
with the full functional, while the non-local parts of the Coulomb
interaction are captured on the DFT level.

The second approximation of our theory is to replace the correction
term in Eq.~\ref{eq:approx1a} by a sum of local correction terms
\begin{eqnarray}
F^{\hat{W}}\approx F_{DFT}^{\hat{W}}
+\sum_R\left(F^{\hat{W}_R}-F_{DFT}^{\hat{W}_R}\right)\,.
\label{eq:approx1}
\end{eqnarray}
In this approximation, the environment of each cluster is replaced by
a non-interacting electron gas. Each term $F^{\hat{W}_R}$ has
precisely the structure of a so-called multi-orbital single-impurity
Anderson model.\cite{anderson:61}

For the sake of completeness, let us quote the error term for the
approximation of Eqs.~\ref{eq:approx1a} and \ref{eq:approx1},
\begin{eqnarray}
\Delta F^{\hat{W}}
&=&
\left(F^{\hat{W}}-\sum_RF^{\hat{W}_R}\right)\nonumber\\
&&-\left(F_{DFT}^{\hat{W}}-\sum_RF_{DFT}^{\hat{W}_R}\right)\,.
\end{eqnarray}

With Eq.~\ref{eq:approx1} we obtain, after
including the particle-number constraint, the following expression for
the total energy.
\begin{widetext}
\begin{eqnarray}
  \bar{E}[\hat{V}_{ext},N]&=&\min_{\{\psi_n,x_n\}}\Biggl\lbrace
  \sum_{n}f(x_n)\langle\psi_n|\frac{\hat{p}^2}{2m_e}+
  \hat{V}_{ext}|\psi_n\rangle
  +\frac{1}{2}\int d^3r
  \,d^3r'\; 
  \frac{e^2 n(\vec{r}) n(\vec{r'})}{4\pi\epsilon_0|\vec{r}-\vec{r'}|}
  +E_{xc}[n(\vec{r})]
  \nonumber\\
  &&+
  \sum_R \Bigl\lbrace
  F^{\hat{W}_R}
\biggl[\langle\pi_\alpha|\psi_n\rangle f(x_n)\langle\psi_n|\pi_\beta\rangle\biggr]
 -\frac{1}{2}\int d^3r
 \, d^3r'\; 
  \frac{e^2 n_R^\chi(\vec{r}) n_R^\chi(\vec{r'})}{4\pi\epsilon_0|\vec{r}-\vec{r'}|}
  -E^{\hat{W}_R}_{xc}[n^\chi(\vec{r})]\Bigr\rbrace
  \nonumber\\
  &&
  -\sum_{n,m}\Lambda_{n,m}\Bigl(\langle\psi_m|\psi_n\rangle-\delta_{n,m}\Bigr)
  -\mu\Bigl(\sum_n f(x_n)-N\Bigr)\Biggr\rbrace\,,
\label{eq:dmftreferencesystemanderson}
\end{eqnarray}
\end{widetext}
where, with $f_n=f(x_n)$,
\begin{eqnarray}
n(\vec{r})&=&\sum_\sigma\sum_n \langle\vec{r},\sigma|\psi_n\rangle f_n
\langle\psi_n|\vec{r},\sigma\rangle\;,
\label{eq:deffulldensity}
\\
n^\chi(\vec{r})&=&\sum_\sigma
\sum_{\alpha,\beta}
\langle\vec{r},\sigma|\chi_\alpha\rangle
\nonumber\\
&&\times\sum_n \langle\pi_\alpha|\psi_n\rangle f_n
\langle\psi_n|\pi_\beta\rangle\langle\chi_\beta|\vec{r},\sigma\rangle\;,
\label{eq:defdensityprojected}
\end{eqnarray}
and
\begin{eqnarray}
n^\chi_R(\vec{r})&=&\sum_\sigma
\sum_{\alpha,\beta\in\mathcal{C}_R}
\langle\vec{r},\sigma|\chi_\alpha\rangle
\nonumber\\
&&\times\sum_n \langle\pi_\alpha|\psi_n\rangle f_n
\langle\psi_n|\pi_\beta\rangle\langle\chi_\beta|\vec{r},\sigma\rangle\,.
\label{eq:deflocaldensity}
\end{eqnarray}
For the sake of simplicity, we write the expression for density
functionals and not for spin-density functionals. The generalization
is straightforward.

The second line in Eq.~\ref{eq:dmftreferencesystemanderson} can be
considered as a correlation correction to the density functional: In the
absence of the correlation correction, the theory recovers the
conventional expression of density-functional theory.

\subsection{Exchange-correlation with local interaction}
\label{sec:xchole}
Finding an expression of the exchange-correlation functional for the
interaction is not straightforward. 

The exchange-correlation energy can be expressed by the
interaction-strength averaged hole function\cite{harris74_jpf4_1170}
$h_\lambda(\vec{r},\vec{r'})$ as
\begin{eqnarray}
E_{xc}[n]\hspace{-1mm}=\hspace{-1mm}
\int d^3r\; n(\vec{r})
\left[\int_0^1 d\lambda\;
\frac{1}{2}\int d^3r'\;\frac{e^2h_\lambda(\vec{r},\vec{r'})}
{4\pi\epsilon_0|\vec{r}-\vec{r'}|}\right]\;.
\label{eq:xcenergyfromhole}
\end{eqnarray}
The hole function is defined via the two-particle density
$n^{(2)}_\lambda(\vec{r},\vec{r'})$ of the ground state for a given
density $n(\vec{r})$ or spin density $n_{\sigma,\sigma'}(\vec{r})$ as
\begin{eqnarray}
h_\lambda(\vec{r},\vec{r'})=
\frac{1}{n(\vec{r})}\left[
n^{(2)}_\lambda(\vec{r},\vec{r'})-n(\vec{r})n(\vec{r'})\right]\;.
\end{eqnarray}
The parameter $\lambda$ scales the interaction $\hat{W}$. The
corresponding hole function is obtained from the wave function for
this scaled interaction.  The $\lambda$-average accounts for the
difference of the kinetic energy between the interacting and the
non-interacting electron gas.  

Because the restricted interaction $\hat{W}_R$ is nonlocal in the
coordinates of each particle, it needs to be modified before it can be
used in a density functional. We propose to use the following model
for the local interaction 
\begin{eqnarray}
v^{\hat{W}_R}(\vec{r},\vec{r'})=
\frac{n^\chi_R(\vec{r})}{n^\chi(\vec{r})}
\frac{e^2}{4\pi\epsilon_0|\vec{r}-\vec{r'}|}
\frac{n^\chi_R(\vec{r'})}{n^\chi(\vec{r'})}\,,
\label{eq:dftmodellocalinteraction}
\end{eqnarray}
where $n_R^\chi(\vec{r})$ and $n^\chi(\vec{r})$ are defined by
Eqs.~\ref{eq:defdensityprojected} and \ref{eq:deflocaldensity}.  This
choice for the interaction is consistent with the expression for the
Hartree energy. It can systematically be extended to the full exchange
correlation energy by including the remaining pair terms.

If we use the model for the restricted interaction from
Eq.~\ref{eq:dftmodellocalinteraction} in
Eq.~\ref{eq:xcenergyfromhole}, we obtain
\begin{eqnarray}
E^{\hat{W}_R}_{xc}=
\int d^3r\; n^\chi_R(\vec{r})
\left[\frac{1}{2}\int d^3r'\;\frac{e^2h(\vec{r},\vec{r'})}
{4\pi\epsilon_0|\vec{r}-\vec{r'}|}
\frac{n^\chi_R(\vec{r'})}{n^\chi(\vec{r'})}
\right]\,.
\label{eq:excrestricedinteraction}
\end{eqnarray}
Here, the $\lambda$-averaged hole function is used.  Hole functions
for some of the commonly used density functionals have been
developed\cite{bahmann08_jcp128_234104,perdew96_prb54_16533}.

Eq.~\ref{eq:excrestricedinteraction} reflects that only electrons on
the same impurity interact with each other. Our approximation assumes
that the exchange-correlation hole of the Anderson impurity model is
identical to that of the fully interacting system. This is strictly
valid for the exchange part, while it is an approximation for the
correlation contribution.

If the hole function is not available, the most simple approximation is
\begin{eqnarray}
E_{xc}^{\hat{W}_R}=
\int d^3r\; n^\chi_R(\vec{r})\epsilon_{xc}[n^\chi_{\sigma,\sigma'}(\vec{r})]
\frac{n^\chi_R(\vec{r})}{n^\chi(\vec{r})}\,,
\end{eqnarray}
where $\epsilon_{xc}$ is the exchange correlation energy per electron.

\subsection{Coulomb matrix elements}
\begin{figure}[h!]
\begin{center}
\includegraphics[width=0.7\linewidth,clip=true]{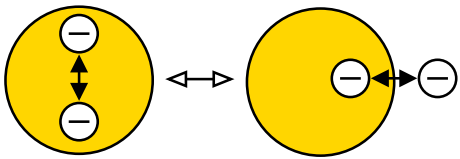}
\end{center}
\caption{\label{fig:inout}Scheme to demonstrate the case where an
  electron leaves a correlated region. In the local approximation the
  interaction with the electron remaining on the correlated cluster is
  completely removed, explaining the need for renormalizing the
  Coulomb interaction.(Color online)}
\end{figure}

One of the important ingredients to our theory is the Coulomb tensor
(\ref{eq:local_interaction}) projected onto the impurity orbitals. The
first idea is to calculate this tensor using the local orbitals
$|\chi_\alpha\rangle$ with the bare Coulomb interaction.

However, as in other local approximations, such an approach would
result in too large Coulomb parameters for the reason schematically
depicted in Fig.~\ref{fig:inout}: If an electron leaves a correlated
orbital, it is likely to be located nearby, so that only a portion of
the interaction is lost. In the local approximation the Coulomb
interaction is completely removed for a particle leaving the
correlated impurity. Thus the energy required to add or remove an
electron from the impurity is overestimated in the local
approximation, respectively charge fluctuations are artificially
suppressed. Therefore, our method will require an appropriate
renormalization of the local U-tensor to account for those screening
channels.

There are several possible approaches to obtain this effect. One can
in principle enlarge the impurity and include more extended orbitals
in the correlated cluster. This will lead at least partially to the
desired renormalization: The  screening is taken over by the extended
orbitals.  Another route, which has proven to be more efficient, is to
perform additional calculations using constrained DFT
\cite{dederichs84_prl53_2512}, which will also provide an estimate of
the renormalization due to screening from other orbitals.

\subsection{Extension to ensemble density-matrix functional theory}

The formulation of the density-matrix-functional theory used so far is
based on pure states, that, however, are not necessarily eigenstates
of the particle-number operator. Usually, density-matrix-functional
theory is formulated for ensembles of many-particle states. The
generalization to ensemble rDMFT is straightforward. A
density-matrix-functional for an ensemble of many-particle states has
the form
\begin{eqnarray}
F^{\hat{W}}({\bm\rho})&=&\min_{\{|\Phi_i\rangle,P_i\}}
\Bigl[
\sum_i P_i \langle\Phi_i|\hat{W}|\Phi_i\rangle
\nonumber\\
&+&\sum_{\alpha,\beta}{h}_{\alpha,\beta}
\left(\sum_iP_i
\langle\Phi_i|\hat{c}^\dagger_\alpha\hat{c}_\beta|\Phi_i\rangle-\rho_{\beta,\alpha}\right)
\nonumber\\
&-&\sum_{i,j}\Lambda_{i,j}\left(\langle\Phi_i|\Phi_j\rangle-\delta_{i,j}\right)
+\lambda\biggl(\sum_iP_i-1)
\Bigr]\,,
\nonumber\\
\end{eqnarray}
where $P_i$ are probabilities for the many-particle states
$|\Phi_i\rangle$. In order to ensure that the probabilities fall
between zero and one, we represent them as square $P_i=X_i^2$ of a
real number $X_i$.

Formally, the theory can be extended to finite temperatures by adding
an additional entropy term $-ST=k_BT\sum_i P_i\ln(P_i)$, that
represents a heat bath, to the density matrix functional.

\section{Evaluation of  the density-matrix functional}
\label{sec:ci}
The reduced-density-matrix functional $F^{\hat{W}_R}$ in
Eq.~\ref{eq:dmftreferencesystemanderson} is determined directly using
the constrained-search formalism of Levy\cite{levy79_pnas76_6062}. The
details of the methodology will be described
elsewhere.

In order to evaluate the density-matrix functional, we translate the
fictitious Lagrangian methodology of Car and Parrinello for ab-initio
molecular dynamics\cite{car85_prl55_2471} to many-particle wave
functions $|\Phi\rangle$.

The problem of ab-initio molecular dynamics and of a many-particle
problem are related, as far as they are applied to the search for
ground states: in both cases one has to determine one or a few of the
lowest states of a highly-dimensional Hamiltonian.  Furthermore, the
fictitious Lagrangian formalism lends itself to the search of a
minimum under constraints. In the Car-Parrinello
method\cite{car85_prl55_2471}, the constraints are the orthonormality
constraints for the Kohn-Sham wave functions. In our case it is the
norm and the requirement that the wave function produces a specified
reduced one-particle density matrix.

We start from a fictitious Lagrangian defined as
\begin{eqnarray}
\mathcal{L}(|\Phi\rangle, |\dot{\Phi}\rangle)
&=&\langle\dot{\Phi}|m_\Phi|\dot{\Phi}\rangle
-\langle\Phi|\hat{W}|\Phi\rangle
\nonumber\\
&-&\sum_{\alpha,\beta}{h}_{\alpha,\beta}
\Bigl(\langle\Phi|\hat{c}^\dagger_\alpha\hat{c}_\beta|\Phi\rangle-\rho_{\beta,\alpha}\Bigr)
\nonumber\\
&+&E\Bigl(\langle\Phi|\Phi\rangle-1\Bigr)\,.
\end{eqnarray}
In practice, each many-particle wave function is represented by a
coefficient array of the Slater determinants. 

Including a friction term, the equation of motion has the form
\begin{eqnarray}
m_\Phi|\ddot{\Phi}\rangle
&=&-(\sum_{\alpha,\beta} {h}_{\alpha,\beta}
\hat{c}^\dagger_\alpha\hat{c}_\beta+\hat{W}-E)|\Phi\rangle
-|\dot{\Phi}\rangle \kappa\,.
\label{eq:fictlageqm}
\end{eqnarray}
This equation of motion is discretized using the Verlet
algorithm\cite{verlet67_pr159_98}, and the constraints are taken into
account using the method of Ryckaert et
al.\cite{ryckaert77_jcompphys23_327}.

The Car-Parrinello method is not only used to optimize the
many-particle wave function for each Anderson impurity model.  It is
also used more conventionally to optimize dynamically the natural
orbitals and occupations in the ``outer loop''. This is analogous to
the conventional Car-Parrinello method, which optimizes the Kohn-Sham
wave functions. Both, natural orbitals and Kohn-Sham wave functions
are single-particle wave functions.

\section{Model calculations}
\label{sec:model}
In order to explore the workings of the formalism described in the
previous sections, we performed calculations on model systems that
allow comparison to exact results.

The model systems are Hubbard chains with $N_s=2$, $N_s=4$, and
$N_s=6$ sites.  Their Hamiltonian is
\begin{eqnarray}
  \hat{H}&=&\hat{T}+\sum_{R=1}^{N_s} \hat{W}_R\,,
\end{eqnarray}
where $\hat{T}$ describes the non-interacting part of the Hamiltonian
\begin{eqnarray}
\hat{T}=-t\sum_{R=1}^{N_s-1}\sum_{\sigma\in\{\uparrow,\downarrow\}}
\biggl(\hat{c}_{R,\sigma}^{\dagger}\hat{c}_{R+1,\sigma}
+\hat{c}_{R+1,\sigma}^{\dagger}\hat{c}_{R,\sigma}
\biggr)\,,
\end{eqnarray}
and $\hat{W}_R$ is the interaction on a the $R$-th site
\begin{eqnarray}
\hat{W}_R=U\hat{n}_{R,\uparrow}\hat{n}_{R,\downarrow}\,.
\end{eqnarray}   
$t$ is the hopping parameter and $U$ is the Coulomb parameter.
$\hat{n}_{R,\sigma}= \hat{c}^\dagger_{R,\sigma}\hat{c}_{R,\sigma}$ is
the particle-number operator for the specified orbital.

We adapted Eq.~\ref{eq:dmftreferencesystemanderson} to the Hubbard
chain by ignoring Hartree and exchange energy terms. They cancel
exactly on this level of the theory.
\begin{widetext}
\begin{eqnarray}
\bar{E}(N)&=&\min_{\{|\psi_n\rangle,x_n\}}\biggl\lbrace
\sum_n f(x_n) \langle\psi_n|\hat{T}|\psi_n\rangle
+\sum_R F^{\hat{W}_R}(\{\rho_{\alpha,\beta}\})
\nonumber\\
&&
-\sum_{n,m}\Lambda_{n,m}\biggl(\langle\psi_m|\psi_n\rangle-\delta_{m,n}\biggr)
-\mu\biggl(\sum_n f(x_n)-N\biggr)\,,
\label{eq:modelenergy}
\end{eqnarray}
where the $f(x_n)$ are the occupations defined in
Eq.\ (\ref{eq:energyfromdmf}),
and the one-particle states
$|\psi_n\rangle$ are the natural orbitals. The density-matrix
functional for for each site has the form
\begin{eqnarray}
F^{\hat{W}_R}\left(\left\{\rho_{\alpha,\beta}\right\}\right)=\min_{|\Phi\rangle}
\biggl[\langle\Phi|\hat{W}_R|\Phi\rangle+\sum_{\alpha,\beta}
h_{\alpha,\beta}\biggl(\langle\Phi|\hat{c}^\dagger_{\alpha}\hat{c}_{\beta}|\Phi\rangle
-\rho_{\beta,\alpha}
\biggr)
-E\biggl(\langle\Phi|\Phi\rangle -1\biggr)\biggr]\,.
\end{eqnarray}
\end{widetext}

%
\begin{figure}[hbt]
 \begin{center}
   \includegraphics[width=\linewidth,clip=true]{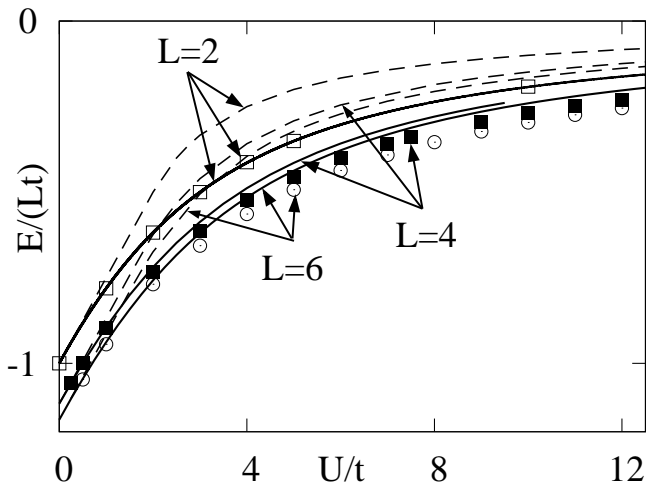}
  \end{center}
  \caption{Energy as function of the interaction strength of the 2-,
    4-, and 6-site Hubbard chains at half filling. Results from our
    theory (symbols) are compared with the exact result (solid line)
    and the mean-field approximation (dashed line).}
  \label{fig:totalenergy}
\end{figure}  
The total energy for half filling is shown in figure
\ref{fig:totalenergy} as function of the interaction strength $U$.  Of
interest is if and how the interaction switches off the covalent
interaction due to left-right
correlation\cite{kolos60_rmp32_205}. Left-right correlation describes
that two electrons forming a bond avoid being on the same site in
order to lower their Coulomb repulsion and plays a major role in
chemistry, because it is vital for understanding the dissociation of
chemical bonds.  Due to this effect, the covalent interaction energy
is completely lost as the interaction grows.  To make contact to the
correlation effects in solids, we note that left-right correlation
describes the same physical mechanism that is responsible for the Mott
band gap.

\begin{figure}
 \begin{center}
 \includegraphics[width=\linewidth,clip=true]
   {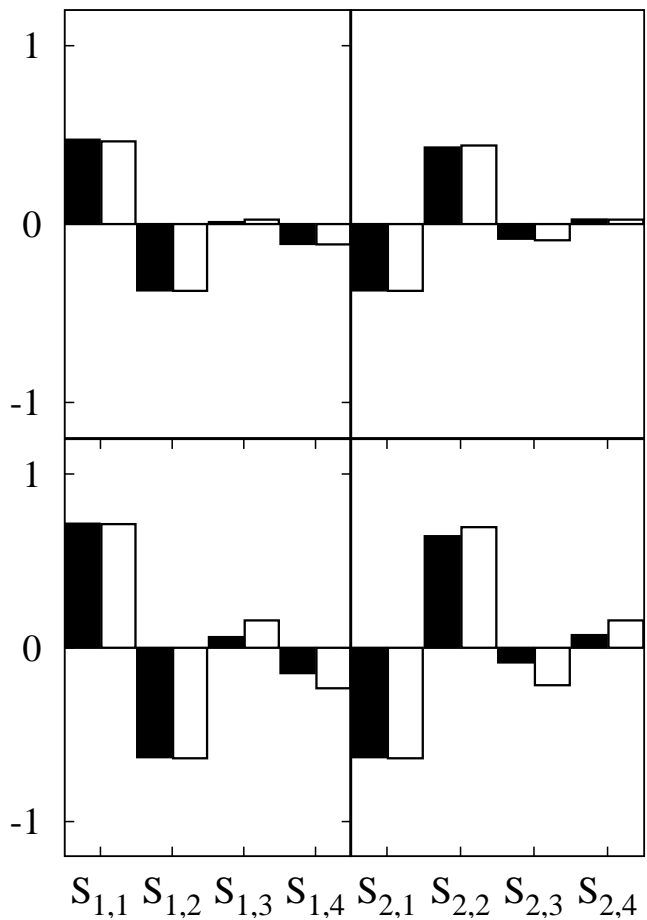}
  \end{center}
  \caption{\label{ANTIFERRO}Spin-spin correlation function
    $S_{R,R'}=\langle\Phi_R|\hat{\vec{S}}_{R}\hat{\vec{S}}_{R'}|\Phi_R\rangle$
    for the 4-site chain with U/t=1 (top) and U/t=8 (bottom) at half
    filling.  The many-particle wave functions $|\Phi_R\rangle$ is
    that of the Anderson model with the impurity at the site of the
    first site index of $S_{R,R'}$. Site one is the left terminal site
    and site two is the left central site. The results from our
    theory, shown as black bars are compared with the exact result for
    the Hubbard chain, shown as white bars.  The results show the
    antiferromagnetic coupling of the approximate ground states. The
    antiferromagnetic correlation grows with increasing $U$. The
    correlation length of the is well described by our theory, even
    though underestimated for large $U$.}
    \label{fig:AFcorrelations}
\end{figure}

For the dimer at half filling, our theory reproduces the exact result,
because the many-particle wave function for the two-particle state is
completely determined by the density matrix.  For the other systems,
our theory underestimates the exact result. This finding is explained
in the appendix~\ref{app:underestimate}.  The underestimation is
strongest for intermediate interactions and vanishes for weak and
for strong interactions. However, most importantly, our theory describes the
nature of the ground state correctly as a
quantum mechanical singlet with proper left-right correlations, which
manifest themselves as strong anti-ferromagnetic correlations in the
present model. This feature is evident from Fig.\
\ref{fig:AFcorrelations}
where we present the spin-spin correlation function
$S_{R,R'}=\langle\Phi_R|\hat{\vec{S}}_{R}\hat{\vec{S}}_{R'}|\Phi_R\rangle$
for the 4-site chain with U/t=1 (top) and U/t=8 (bottom) at half
filling compared to the exact result. Note that one observes an
increase of the nearest-neighbor correlation with increasing $U$,
which is in accordance with the general tendency of the Hubbard model
towards anti-ferromagnetism at half filling.  The
  correlation length is well reproduced within our theory, albeit
  somewhat underestimated as compared to the exact result. The latter
  point is to be expected, because the one-dimensional Hubbard model
  is a Luttinger liquid with a slow algebraic decay of spin-spin
  correlations, while for our local approximation we can expect that
  the corresponding correlation function is rather that of a Fermi
  liquid with the standard Friedel oscillation exponent.

A correct description of the ground state, in particular reproducing
the proper anti-ferromagnetic correlations, is by no means
trivial. The reason is that left-right correlation in a singlet cannot
be captured by a single Slater determinant. Thus, it cannot be
obtained in an independent-electron picture. 

In order to compare our results with the independent-electron picture,
we performed calculations based on a static mean-field theory.  To
generate a static mean-field description within our density-matrix
functional framework, one simply has to replace $F^{\hat{W}_R}[\rho]$
in Eq.~\ref{eq:modelenergy} by the expression
$W_R^{MF}[\rho]=U\rho_{R\uparrow,R\uparrow}\rho_{R\downarrow,R\downarrow}$.

The ground state in the mean-field approximation is a single
Slater determinant.  For small interaction strengths, the wave function is
formed from the two bonding orbitals and the electrons remain
uncorrelated. As a consequence the total energy raises with constant
slope with the interaction strength. At approximately $U=2t$, the
mean-field ground-state wave function undergoes a transition to an
anti-ferromagnetic state.  As in a typical second order transition,
the magnetization grows with an approximate square-root behavior away
from the transition. In the large-U limit, each site has the magnetic
moment of one electron. The anti-ferromagnetic state is a so-called
broken-symmetry state, which is not an eigenstate of the total spin.
For the dimer, such a broken-symmetry state can be described as a
superposition of a singlet and a triplet state. The true ground state, however,
is a pure singlet, because only the singlet state can profit from some
remaining covalency. 

Another interesting quantity for the Hubbard model is the double
occupancy, defined as
$d_R=\langle\Phi|\hat{n}_{R,\uparrow}\hat{n}_{R,\downarrow}|\Phi\rangle$.
In the Hubbard model, the interaction energy depends exclusively on
the sum of the double occupancies.

In our theory, we extract the double occupancy of a given site from
the Anderson impurity model with the interaction at that site. Hence,
the double occupancy is calculated for each site from a different
Anderson impurity model and thus from a different wave function.

\begin{figure}[h!]
 \begin{center}
   \includegraphics[angle=-90,clip=true,width=\linewidth]
   {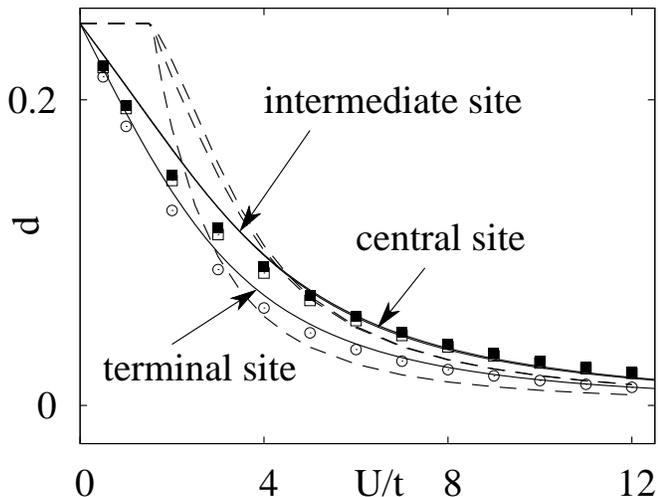}
  \end{center}
  \caption{Double occupancy the 6-site Hubbard chain at half-filling
    as function of the interaction strength from our theory (symbols)
    in comparison with the exact result (lines). Finite-size effects
    can be estimated by comparing the central sites (black squares)
    with the subsurface (open squares) and surface sites (open
    circles). The exact result does not distinguish appreciably
    between central and subsurface sites, which indicates that the
    underestimation by our theory is not a finite size effect.}
  \label{fig:DoubleOccupancy}
\end{figure}  

The double occupancy for the 6-site Hubbard chain is shown in figure
\ref{fig:DoubleOccupancy}. The continuous suppression of the double
occupancy with increasing interaction strength is well reproduced.
Also the difference between surface sites and the center of the chain
is in agreement with the exact calculation. However, our theory tends
to underestimate the double occupancy resulting in a somewhat steeper
slope in the small-U limit. The large-U limit is well reproduced.

The double occupancy provides clear evidence for the strength of the
presented theory as compared to the mean-field theory.  In the
mean-field theory, the double occupancy remains at $d=\frac{1}{4}$
until $U\approx2t$, from where it starts to decrease as the system
builds up anti-ferromagnetic correlations. In the exact result, and in
our approximation, the double occupancy starts to decrease as soon as
the interaction is switched on. The failure of mean-field theory is
due to its inability to produce the correct singlet ground state.

While the Anderson impurity models provide a local perspective and
insight on electron correlations as they are present in the
many-particle wave function, the density matrix provides the global
perspective. 
The density matrix is fully described by a set of one-particle
orbitals, the natural orbitals, and their occupations.  While the
natural orbitals are distinct from Kohn-Sham orbitals of
density-functional theory, they provide similar type of information on
the system.

\begin{figure}
 \begin{center}
  \includegraphics[clip=true,width=\linewidth]{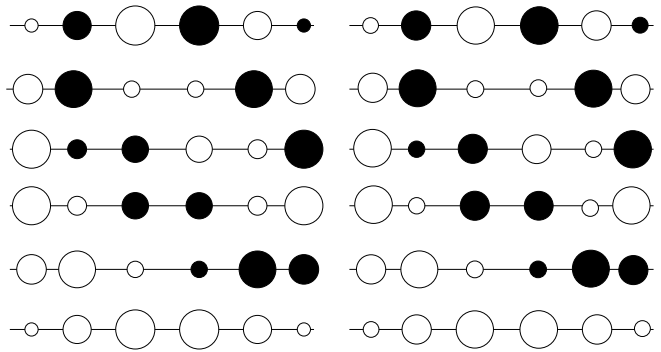}
  \end{center}
  \caption{\label{fig:Natural_orbitals}Natural orbitals of the 6-site
    Hubbard chain for U=0 (left) and U/t=8 (right). The radius of each
    sphere is proportional to the amplitude of the natural orbital,
    and the area is proportional to the site occupation. Full and open
    circles represent the different signs of the orbital
    coefficients. The orbitals are arranged with decreasing occupation
    from bottom to top. For each spatial orbital shown, there are two
    spin orbitals.}
  \end{figure}

\begin{figure}
 \begin{center}
   \includegraphics[angle=-90,width=\linewidth,clip=true]
   {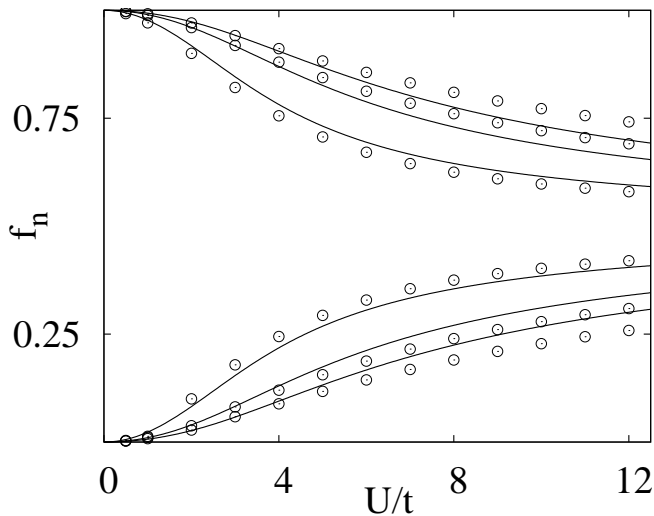}
  \end{center}
  \caption{\label{fig:EIGENVALUES_RHO_IJ}Occupations $f_n$ of the
    natural orbitals as function of the interaction strength from our
    theory (circles) compared to exact result. The occupations become
    more equally distributed with increasing interaction strength.}
\end{figure}

It is striking that the natural orbitals depend very little on the
interaction strength as seen in figure~\ref{fig:Natural_orbitals}.
Qualitatively there is no difference and the quantitative changes are
barely visible in this representation.
  However, the interaction strength has a strong effect on the
  occupations as shown in figure~\ref{fig:EIGENVALUES_RHO_IJ}. While
  the occupations in the (non-degenerate) ground state of a
  non-interacting system are always either zero or one, all
  occupations become fractional if the interaction is switched on.
  From a chemical point of view, the depopulation of bonding orbitals
  and the population of formerly unoccupied antibonding orbitals
  reflects the depression of the covalent interaction by the
  interaction.  From a solid-state point of view, the behavior seen in
  Fig.~\ref{fig:EIGENVALUES_RHO_IJ} is reminiscent of the momentum
  distribution functions of a Fermi-liquid or the Luttinger
  liquid. While for noninteracting particles the momentum distribution
  has a sharp drop of the occupations from one to zero at the Fermi
  momentum, for the interacting electron systemm, it exhibits a
  gradual decrease with either a drop of size $Z^{-1}<1$, the
  quasi-particle renormalization factor of a Fermi liquid, or, for a
  Luttinger liquid, a sharp change with an algebraic singularity at
  the Fermi momentum

The smearing out of the occupations implies that the number of
one-particle wave functions to be considered must be larger than that
required for density-functional theory. If we venture into an
extrapolation of our findings to real systems, we expect that it will
be necessary to include one-particle wave functions $|\psi_n\rangle$ up
to the antibonding orbitals of bonds for which left-right correlation
is important. This implies that we need to include orbitals up to
several electron volts above the Fermi-level. Working in our favor is
that in the strongest bonds, with high-lying antibonding states, the
kinetic energy is large and therefore dominating over the Coulomb
repulsion.  Correlations are dominant in relatively weak bonds, either
when they are stretched to the dissociation limit, or when they
involve d- and f-electron orbitals or $\pi$ bonds.

One of the major problems plaguing density-functional calculations is
the band-gap problem. Here, we do not refer to the optical band gap,
which is an excited state property and which, in a strict sense, is
neither accessible by ground-state DFT nor by our theory. Instead we
investigate here the thermodynamic gap, which we define as
\begin{eqnarray*}
\Delta=\lim_{\delta\rightarrow0}\biggl(
\left.\frac{dE}{dN}\right|_{N+\delta}
-\left.\frac{dE}{dN}\right|_{N-\delta}\biggr)\,.
\end{eqnarray*}
Because $E(N)$ is a series of line-segments interpolating between the
values for integer particle numbers $N$, this definition is identical
to the more common expression $\Delta=E(N+1)-2E(N)+E(N-1)$, which is
the difference $\Delta=A-I$ between electron affinity $A$ and
ionization potential $I$.  Because $E(N)$ is a true ground-state
property, it should be properly described both by our theory and by
correct density-functional theory. A failure of describing the
thermodynamic gap has a fundamental impact on the defect chemistry in
semiconductors and insulators.
\begin{figure}
 \begin{center}
   \includegraphics[width=\linewidth,clip=true]
   {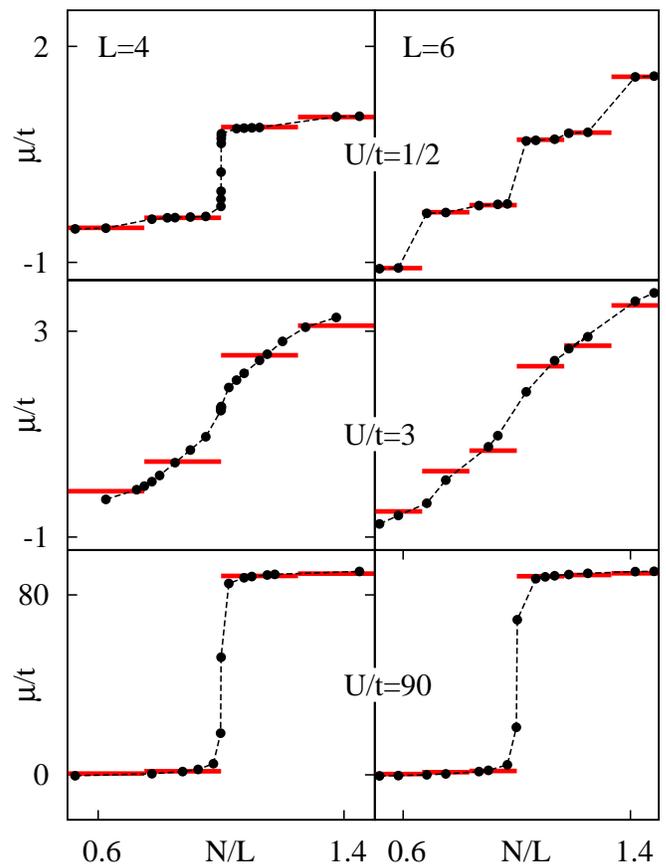}
  \end{center}
  \caption{Chemical potentials $\mu$ versus filling $N/L$ from our
    theory (spheres and dashed lines) and the exact Hubbard chain
    (solid lines) for the four-site (left) and the six-site (right)
    Hubbard chains as function of the particle number for $U/t=1/2$,
    $U/t=3$ and $U/t=90$ from top to bottom. The dashed lines are a
    guide to the eye and connect the data points. (Color online)}
  \label{fig:Chemicalpotential}
\end{figure}  

The chemical potential $\mu(N)=\frac{dE}{dN}$ for different
interaction strengths calculated with our theory is shown in
figure~\ref{fig:Chemicalpotential} and compared to the exact
thermodynamic band gap, included as horizontal lines in the figure. It
is obvious that our theory does not yield a discontinuity. Despite of
this, the value of the band band gap can nevertheless be obtained
using Slater's transition rule as
$\Delta=\mu(N+\frac{1}{2})-\mu(N-\frac{1}{2})$ from the difference of
the chemical potentials at half-integer occupations.
\begin{figure}
 \begin{center}
   \includegraphics[width=\linewidth,clip=true]{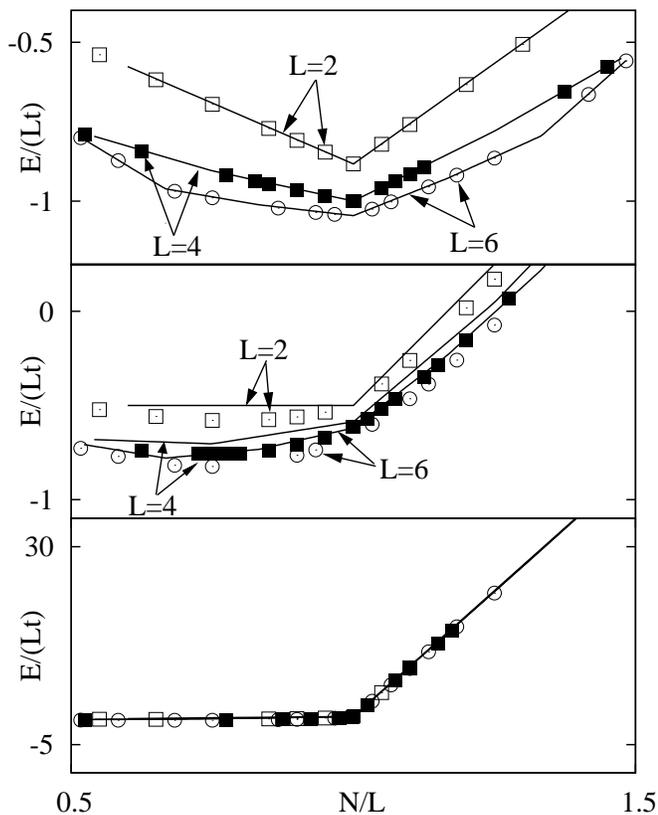}
  \end{center}
  \caption{Total energy for fractional particle numbers N divided by
    the number of sites L of the Hubbard chains for $U/t=1/2$ (first
    panel), $U/t=3$ (second) and $U/t=90$ (third). Symbols denote results from our theory, lines the exact behavior.}
  \label{ENL}
\end{figure}  

The failure to describe the discontinuity of the chemical potential,
and thus the band gap, reminds of the behavior of approximate
density-matrix functionals.\cite{helbig07_epl77_67003} These
functionals do not make the local approximation, but use approximate
numerical representations of the exchange correlation energy. Also
here, the discontinuity vanishes, while the band gap calculated with
half-occupied orbitals are satisfactory predictions.

Except for the non-interacting case, the total energy appears to be a
continuous function of the particle number within our theory, while
the exact result has jumps in the slope at integer fillings.  We
attribute this deficiency of our theory to its tendency to
underestimate the total energy. As seen in figure~\ref{ENL} this
underestimation is most prominent for fractional particle numbers at
intermediate interaction strengths. For small or huge values of $U$
the effect is less visible; but here, too, the variation across
integer fillings is smooth, as it is apparent from Fig.\
\ref{fig:Chemicalpotential}. As discussed in
appendix~\ref{app:underestimate} this defect is related to the fact
that the wave function for each Anderson-impurity model is optimized
individually. An approximation based on a common wave function for all
sites may direct towards a remedy of this problem.

\section{Conclusion}
We proposed a pathway to include explicit correlations into
density-functional based calculations. The guiding ideas have been to
develop a strict variational approximation of the interacting electron
gas, that avoids the dimensional bottleneck of may-particle quantum
theory, and that allows seamless incorporation into electronic
structure methodology of density-functional theory.

Let us summarize what we have accomplished so far: Based on a general
formulation using a density-matrix-functional description of the
solid, the many-particle problem for the lattice with the full
interaction has been replaced by a collection of generalized
multi-orbital Anderson impurity models, defined through the
functionals $F^{W_R}$, see Eq.~(\ref{eq:approx1}).
These Anderson impurity models are defined relative to the DFT
functional as accurate way to obtain the quasi-particle dispersion
including effects of non-local correlations and proper
screening. Since the local DFT contributions can be explicitly
subtracted from the density-matrix functional, the problem of double
counting can be treated without further assumptions in this approach.
Restricting the actual correlations to one site only has proven to be
a powerful idea in many-particle schemes such as Dynamical Mean-Field
theory.  We believe that the density-matrix
  functional approach plus local approximation as proposed in this
  paper, when properly implemented, will provide a tool to treat
  correlation effects in a spirit similar to DMFT, however using the
information of the density-matrix only.

As test case we applied our method to finite Hubbard chains as simple
model systems. The extremalization of the the density-matrix
functional is done on the fly using an explicit constrained-search
algorithm\cite{levy79_pnas76_6062}. This iterative approach has been
solved by a dynamical approach inspired by the Car-Parrinello
method. The iterative nature of the constrained-search algorithm holds
promise to important cost reductions in an outer self-consistency
loop.

In contrast to static mean-field theory, our method produces at half
filling the correct singlet state with strong anti-ferromagnetic
correlations, and thus gives a qualitative description of left-right
correlation. The local approximation smears out the band-gap, which
has been observed similarly in approximate density-matrix
functionals. The energy levels of the many-particle system are,
however, well approximated using the chemical potentials half-integer
particle numbers.

Presently, each of these Anderson impurity models still represents a
nontrivial many-particle problem on the infinite lattice. Thus,
apparently nothing has been gained so far regarding a calculation of
physical properties in this approximation. The important next step now
is to reformulate the effective quantum impurity problem in a way that
it can be solved efficiently. In standard DMFT the important aspect is
that the actual lattice structure only enters through the density of
states. As consequence, the solution of the effective quantum impurity
problem can be done without ever resorting to the actual lattice.

It is obviously important to develop a similar approach for the local
density-matrix functional. The most straightforward idea here would be
to treat the quantum impurity as open quantum system and apply
techniques developed in this field to facilitate the creation of an
efficient impurity solver for our theory.

\begin{acknowledgements}
We thank Michael Pothoff and Karsten Held for stimulating
discussions. Financial support by the Deutsche Forschungsgemeinschaft
through SPP1145 and FOR1346 is gratefully acknowledged.
\end{acknowledgements}

\appendix
\section{Local approximation}
\label{app:underestimate}
Here we show that distributing the interaction into many
Anderson-impurity models leads to an underestimation of the energy,
i.e.
\begin{eqnarray}
F^{\sum_R \hat{W}_R}[{\bm\rho}]\ge\sum_R F^{\hat{W}_R}[{\bm\rho}]\,.
\label{eq:underestimatelocalappr}
\end{eqnarray}

Let $|\Phi_0[{\bm\rho}]\rangle$ be the ground-state wave function for the
interaction $\sum_R \hat{W}_R$ and a one-particle term $\hat{h}_0$ so that
\begin{eqnarray}
  \left[\hat{h}_0[{\bm\rho}]
+\sum_R\hat{W}_R-\mathcal{E}_0[{\bm\rho}]\right]|\Phi_0[{\bm\rho}]\rangle=0\,.
\end{eqnarray}
The one-particle term $\hat{h}_0[{\bm\rho}]$ is determined such that
$|\Phi_0[{\bm\rho}]\rangle$ produces the specified reduced density
matrix $\bm{\rho}$.  Let furthermore $|\Phi_R[{\bm\rho}]\rangle$ be the
ground-state wave function with only a local interaction on site $R$ and
with the specified reduced density matrix:
\begin{eqnarray}
  \left[\hat{h}_R[{\bm\rho}]
+\hat{W}_R-\mathcal{E}_R[{\bm\rho}]\right]|\Phi_R[{\bm\rho}]\rangle=0\,.
\end{eqnarray}

With these definitions, we obtain the desired inequality
Eq.~\ref{eq:underestimatelocalappr}:
\begin{eqnarray}
F^{\sum_R\hat{W}_R}[{\bm\rho}]
=\sum_R \biggl\langle\Phi_0[{\bm\rho}]\biggr|\hat{W}_R
\biggl|\Phi_0[{\bm\rho}]\biggr\rangle
\nonumber\\
\ge
\sum_R \biggl\langle\Phi_R[{\bm\rho}]\biggr|\hat{W}_R
\biggl|\Phi_R[{\bm\rho}]\biggr\rangle
=\sum_R F^{\hat{W}_R}[{\bm\rho}]\,.
\end{eqnarray}
The argument rests on the simple fact that the energy decreases, if
one optimizes wave functions for each impurity model individually,
rather than to select one wave function that needs to suit all local
interactions simultaneously.

\end{document}